\documentclass[reprint, prl, superscriptaddress]{revtex4-2}
\usepackage{amsmath,amssymb}
\usepackage{stmaryrd}
\usepackage{latexsym}
\usepackage{CJK}
\usepackage{graphicx}
\usepackage{indentfirst}
\usepackage{listings}
\usepackage{xcolor}
\usepackage{booktabs}
\usepackage{stmaryrd}
\usepackage{multirow}
\usepackage{verbatim}
\usepackage{makecell}
\usepackage{lineno,hyperref}
\usepackage{longtable}
\usepackage{hyperref}
\usepackage{upgreek}
\usepackage{mathtools}
\usepackage{comment}
\usepackage{enumitem}
\usepackage[figuresright]{rotating}
\usepackage{epstopdf}
\usepackage{times}
\usepackage{mathptmx}

\hypersetup{colorlinks,breaklinks,
            linkcolor=blue,urlcolor=blue,anchorcolor=blue,citecolor=blue}

\newcommand{\rmd}{\mathrm{d}}

\newcommand{\usgn}{\mathrm{sgn}}
\newcommand{\utranspose}{^\mathsf{T}}
\newcommand{\brho}{\boldsymbol{\rho}}

\begin{document}
\title{Disconnection Flow-Mediated Grain Rotation}

\author{Caihao Qiu}
\affiliation{Department of Materials Science and Engineering, City University of Hong Kong,  Hong Kong SAR, China}
\author{Marco Salvalaglio}
\affiliation{Institute of Scientific Computing, TU Dresden, 01062 Dresden, Germany}
\affiliation{Dresden Center for Computational Materials Science, TU Dresden, 01062 Dresden, Germany}
\author{David J. Srolovitz}
\affiliation{Department of Mechanical Engineering, The University of Hong Kong, Pokfulam Road, Hong Kong SAR, China}
\author{Jian Han}
\affiliation{Department of Materials Science and Engineering, City University of Hong Kong, Hong Kong SAR, China}


\begin{abstract}
Interface migration in microstructures is mediated by the motion of line defects with step and dislocation character, i.e., disconnections. We propose a continuum model for arbitrarily-curved grain boundaries or heterophase interfaces accounting for disconnections' role in grain rotation. Numerical simulations show that their densities evolve as grain size and shape change, generating stresses and increasing or decreasing misorientation; the predictions agree with molecular dynamics simulations for pure capillarity-driven evolution, and are interpreted through an extended Cahn-Taylor model.
\end{abstract}

\maketitle

Grain rotation is widely observed during microstructure evolution in polycrystals.
It may occur with or without plastic deformation inside the grains.
In the former case, crystal lattice rotation is associated with dislocation motion within grains~\cite{wang2014grain, zhou2017reversal,sal2019closing,li2020shear}.
In the latter case, it is associated with grain boundary (GB) sliding, as observed in experiments~\cite{harris1998grain, tian2022situ} and atomic-scale simulations ~\cite{upmanyu2006simultaneous, trautt2012grain,wu2012phase}.
In this letter, we focus on the mechanisms underlying the GB sliding that produces grain rotation without plastic deformation within grains.

Earlier models proposed that grain rotation is driven by  the reduction in grain boundary energy $\bar{\gamma}$  associated with the dependence of  $\bar{\gamma}$ on the relative misorientation of the grains meeting at the GB, $\theta$
~\cite{harris1998grain, kobayashi2000continuum, richard2015continuous}.
Molecular dynamics (MD) and phase field crystal (PFC) simulations showed that some grains rotate in the sense that increases $\bar{\gamma}$~\cite{haslam2001mechanisms, srinivasan2002science, trautt2012grain, waters2022grain, wu2012phase}.
Observations over the past 70 years demonstrate that GB migration is often accompanied by shear across the GB; i.e., shear-coupled migration~\cite{li1953stress, bainbridge1954recent, cahn2006coupling}.
The coupling factor is defined as $\beta = v_\parallel/v_\perp$, where $v_\parallel$ and $v_\perp$ are the rates of GB shear and migration, respectively, in an unconstrained bicrystal with a nominally flat GB~\cite{ashby1972boundary,cahn2006coupling}.
Cahn and Taylor described the simultaneous change in grain size and grain lattice rotation based on $\beta$~\cite{cahn2004unified, taylor2007shape}.
They implicitly assumed that $\beta$ is independent of  GB inclination. 
We and others~\cite{han2022disconnection,sal2022disconnection,qiu2023interface, joshi2022finite} recently described more complex GB behaviors (e.g., the inclination and/or temperature dependence) based on crystallography-deduced ``effective'' $\beta$’s, without assuming shear-coupled migration factors.

Other studies focused on the underlying mechanism of shear-coupled migration in low-angle GBs (which may be described as an array of dislocations) as lattice dislocation glide~\cite{cahn2006coupling}.
Grain rotation was indeed simulated as the result of lattice-dislocation dynamics~\cite{bobylev2012grain, zhang2018motion, zhang2020new}.
Dislocation Burgers vectors in high-angle GBs may be accounted for using the Frank-Bilby approach~\cite{frank1950resultant,bilby1955types,cahn2006coupling}, i.e., geometrically necessary dislocations (GNDs).
In the approach of Admal et al.~\cite{admal2018unified,he2021polycrystal}, the GND density tensor implies that the  GBs respond to stress.
However, GNDs are not real/physical dislocations, and the evolution of the GND density does not properly describe high-angle GB kinetics.
On the other hand, experiments and simulations have demonstrated that shear-coupled migration is associated with the glide of disconnections --  line defects possessing both dislocation and step character~\cite{bollmann1970general,ashby1972boundary,hirth1973grain,rajabzadeh2014role,zhu2019situ}.
The disconnection description of GB migration/shear coupling works for both the special case of low-angle GBs and all other (including high-angle) GBs~\cite{han2018grain}. However, no model has yet shown their role in grain rotation.
In this paper, we show how disconnection dynamics lead to grain rotation during GB motion.

\begin{figure}[t]
\includegraphics[width=0.8\linewidth]{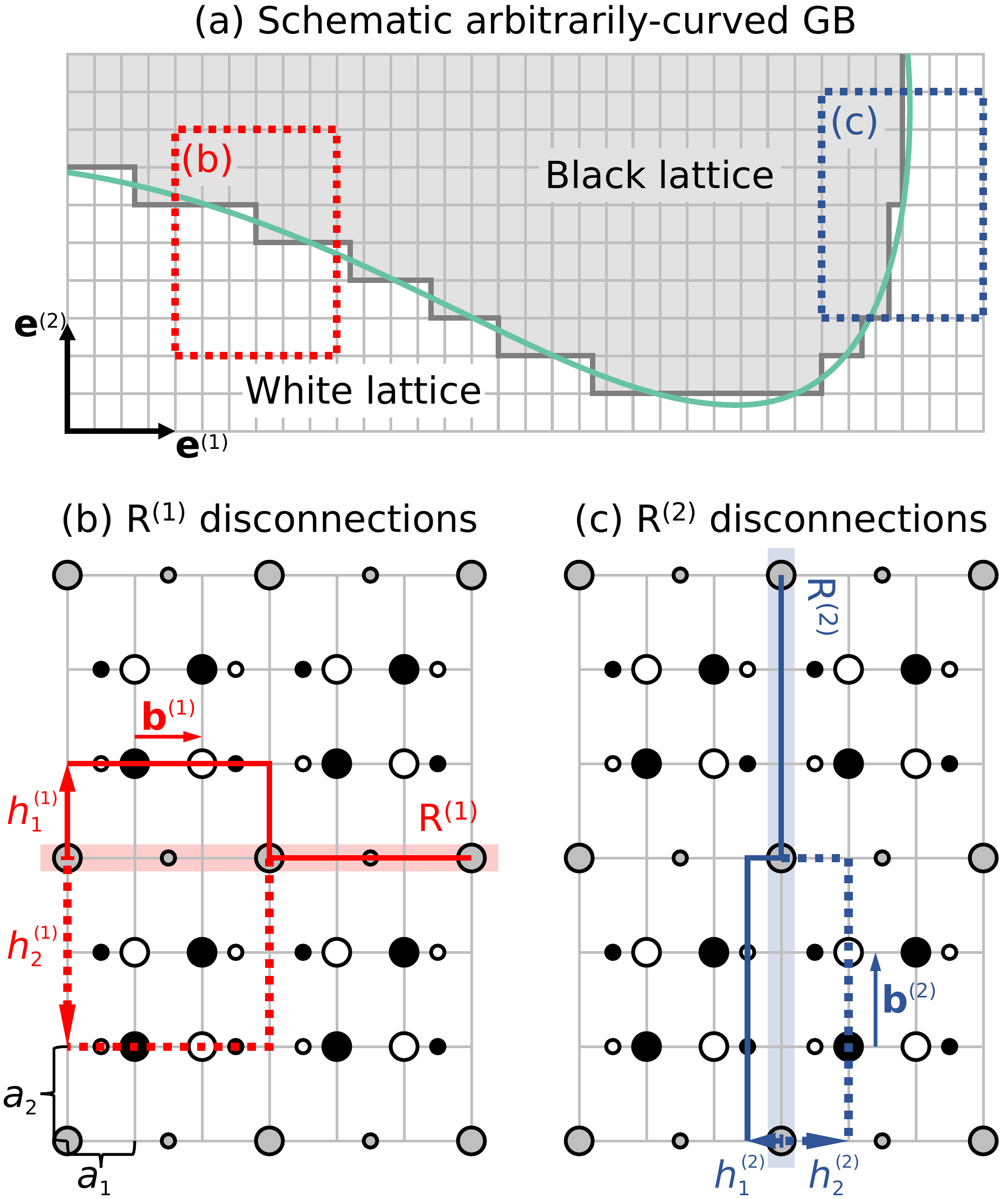}\hspace{-1.78em}%
\caption{(a) The disconnection description of an arbitrarily-curved GB (green curve) where the black lattice and the white lattice meet together.
Dichromatic pattern for a $\Sigma 3$ [110] tilt GB in an FCC crystal.
The gray grid shows the DSC lattice; $a_1$ and $a_2$ are  DSC lattice parameters.
The red and blue lines represent the highly-coherent reference interfaces, R$^{(1)}$ and R$^{(2)}$.
(b) Displacement of the black lattice by the red arrow $\mathbf{b}^{(1)} = (a_1, 0, 0)$ leads to a disconnection with Burgers vector $\mathbf{b}^{(1)}$ and step height $h_1^{(1)}$ or $h_2^{(1)}$.
(c) Relative displacement by the blue arrow $\mathbf{b}^{(2)} = (0, a_2, 0)$ forms a disconnection with Burgers vector $\mathbf{b}^{(2)}$ and step height $h_1^{(2)}$ or $h_2^{(2)}$.
}
\label{fig_dichromatic}
\end{figure}


We represent the GB profile as a continuous curve in two dimensions as $\mathbf{x}(s) = \left(x_1(s), x_2(s)\right)\utranspose$, where $s$ parameterizes the curve (green curve in Fig.~\ref{fig_dichromatic}a).  
The bicrystallography of the GB is described by the superposition of the crystal lattices of the two grains (different orientations) meeting there.
Figure~\ref{fig_dichromatic} shows the example of two face-centered cubic (FCC) lattices (denoted by black and white points) rotated by $70.53^\circ$ about $[110]$; the gray points represent coinciding lattice points and the gray lines indicate the displacement-shift-complete (DSC) lattice~\cite{bollmann1970general, hirth1973grain}.
A flat GB is represented  by choosing a DSC lattice plane and deleting all black lattice points on one side and the white points on the other.
A DSC lattice plane with a high density of points commonly corresponds to a GB with low excess energy~\cite{han2018grain}.
The dichromatic patterns for $[110]$ tilt GBs possess two sets of highly coherent planes; we treat these as reference interfaces R$^{(1)}$ and R$^{(2)}$.
Displacing part of one lattice (e.g., the black lattice) by a DSC vector along R$^{(1)}$, $\mathbf{b}^{(1)}$, generates a disconnection; see Fig.~\ref{fig_dichromatic}b.
Different choices of the GB plane after the displacement of the black lattice implies that multiple disconnection step heights are possible (e.g., $h_1^{(1)}$ and $h_2^{(1)}$); this feature plays a central role in  grain rotation, as shown below.
There are multiple disconnection modes on R$^{(1)}$, denoted as $(\mathbf{b}^{(1)}, h_m^{(1)})$; the subscript $m$ labels the mode and the superscript $(1)$ labels the corresponding reference interface; similarly, R$^{(2)}$ disconnections are $(\mathbf{b}^{(2)}$, $h_m^{(2)})$ (see Fig.~\ref{fig_dichromatic}c).

A continuously-curved GB is described by R$^{(1)}$ and R$^{(2)}$ disconnections, see Fig.~\ref{fig_dichromatic}a; the relationship between  disconnection density and  GB profile is $\sum_m h_m^{(1)} \rho_m^{(1)} = -l_2$ and $\sum_m h_m^{(2)} \rho_m^{(2)} = l_1$, where $\rho_m^{(k)}$ is the density (number per length) of the $m^{\rm th}$ mode of R$^{(k)}$ disconnections and $\mathbf{l} = (l_1, l_2)^\mathsf{T}$ is the local unit tangent vector \cite{han2022disconnection, qiu2023interface}.
Since $\{\rho_m^{(k)}\}$ and $\mathbf{l}$ are not bijectively related, the temporal evolution of both the GB profile (continuum-scale), $\mathbf{x}(t)$, and the disconnection densities (crystallographic-scale), $\{\rho_m^{(k)}(t)\}$, are necessary to describe GB dynamics.

The total free energy of a system with a GB, $E[\mathbf{x}, \{\rho_m^{(k)}\}]$, is a function of the GB free energy, elastic free energy generated by the disconnection Burgers vectors and the  free energies of the grains (see the Supplemental Material, SM, for details).
The driving forces for the evolution of the GB network are the variations of the total free energy with respect to  $\mathbf{x}$ and $\rho_m^{(k)}$.
Assuming  overdamped dynamics, the equation of motion for the GB profile with two reference interfaces is (see SM for detailed derivations)
\begin{equation}\label{EOM_x}
\dot{\mathbf{x}}
=
\mathbf{M}_x \sum_m \left( \Gamma \mathbf{l}^\prime_m - \sigma \boldsymbol{\Lambda}_m + \psi \mathbf{n}\right),
\end{equation}
where $\mathbf{M}_x$ is the mobility tensor \cite{lobkovsky2004grain, han2022disconnection, Chen2020grain}
for GB profile evolution.
$\Gamma\mathbf{l}^\prime_m$ represents the capillarity force, where $\Gamma = \gamma + \gamma_{, \phi\phi}$ is the GB stiffness ($\gamma$ is the GB free energy and $\phi = \arctan(l_2/l_1)$ is the inclination angle) and  $\mathbf{l}^\prime_m = \rmd (- h_m^{(2)} \rho_m^{(2)}, h_m^{(1)} \rho_m^{(1)})\utranspose/\rmd L$ refers to the local curvature expressed as the derivatives of disconnection (step) density.
$\sigma \boldsymbol{\Lambda}_m$ is the elastic driving force, where $\sigma \equiv \mathbf{e}_1 \cdot \boldsymbol{\sigma} \mathbf{e}_2$ ($\boldsymbol{\sigma}$ is the stress tensor arising from the self- and external stress and $\mathbf{e}_k$ is the direction of R$^{(k)}$ interface; see Fig.~\ref{fig_dichromatic}).
$\boldsymbol{\Lambda}_m$ is a term related to  shear-coupled migration (see SM):
\begin{equation}\label{Lamda_m}
\boldsymbol{\Lambda}_m
\equiv
\left(\begin{array}{c}  b_m^{(1)}\rho_m^{(1)} + \dfrac{\usgn(l_2)}{l_1} \left| h_m^{(1)}\rho_m^{(1)}\right|  \displaystyle \sum_m b_m^{(2)}\rho_m^{(2)}
 \\
 b_m^{(2)}\rho_m^{(2)} + \dfrac{\usgn(l_1)}{l_2} \left| h_m^{(2)}\rho_m^{(2)}\right| \displaystyle \sum_m b_m^{(1)}\rho_m^{(1)}
 \end{array}\right).
\end{equation}
$\psi \mathbf{n}$ is the difference in the free energy density of the grains across the GB, where $\mathbf{n} = (-l_2, l_1)\utranspose$ is the unit normal vector.

\begin{figure*}[htpb]
\includegraphics[width=0.9\linewidth]{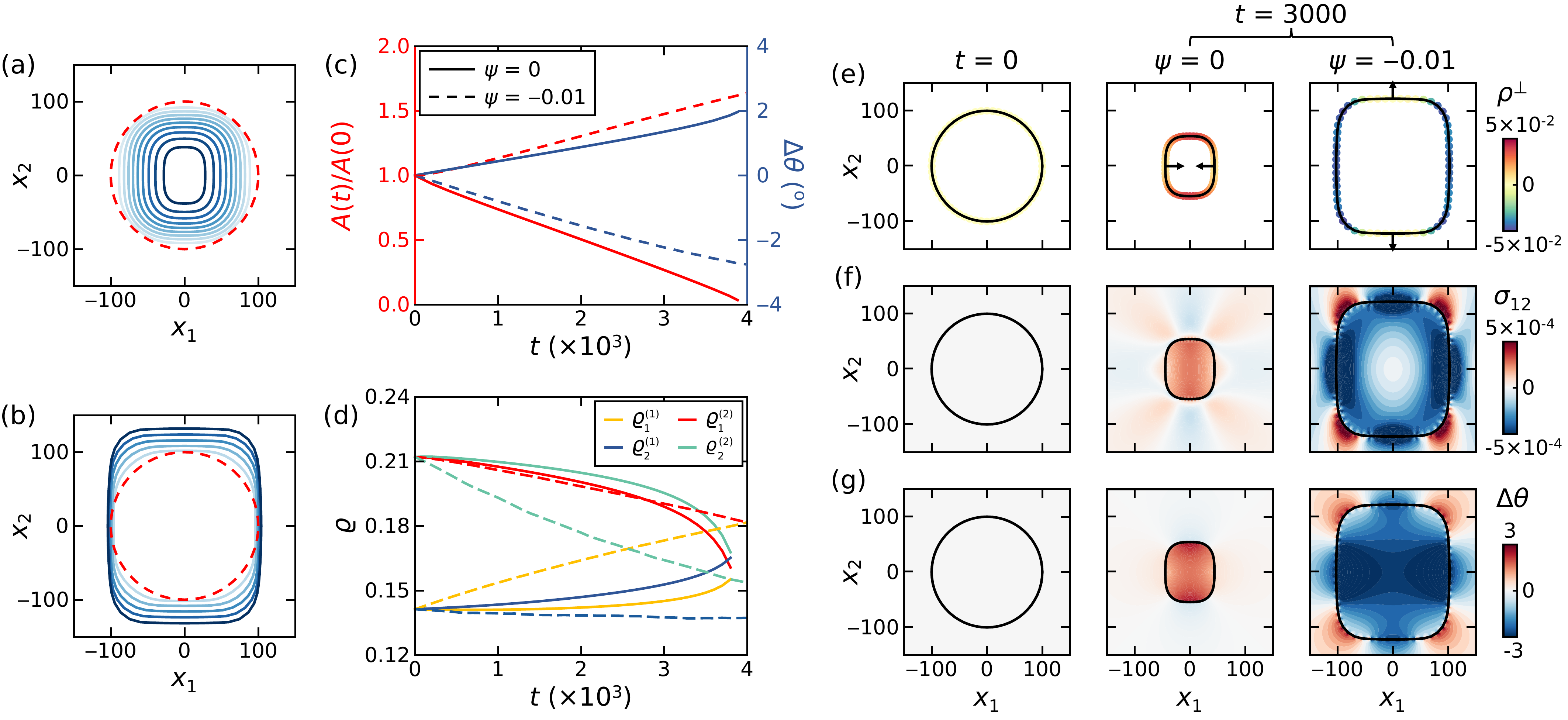}\hspace{-1.78em}%
\caption{(a) Evolution of an initially circular interface considering two disconnection modes. The red dash lines are the initial profiles, and the time interval is $\Delta t$ = 200.
(b) Evolution of an initially circular interface under an extra free energy density jump $\psi = -0.01$, and the time interval is $\Delta t$ = 400.
(c) The normalized domain area $A(t)/A(0)$ vs. time $t$ and the misorientation change at the center of the domains $\Delta \theta$ vs. time $t$.
Solid lines represent the $\psi = 0$ case, and dashed lines refer to the $\psi = -0.01$ case.
(d) Average disconnection densities $\varrho_1^{(1)}, \varrho_1^{(2)}, \varrho_2^{(1)}$, and $\varrho_2^{(2)}$ vs. time $t$. Solid and dashed lines as in panel c.
(e) Disconnection densities perpendicular to the GB at $t = 0$ and the cases with and without free energy density jump at $t =  3000$.
(f) Rotation fields $\Delta \theta$ and
(g) stress fields $\sigma_{12}$ of the corresponding microstructures.}
\label{fig_example}
\end{figure*}

The disconnection density evolution along the GB  is (see SM for details)
\begin{widetext}
\begin{equation}\label{EOM_rho}
\left(\begin{array}{c}
\dot{\brho}^{(1)} \\ \dot{\brho}^{(2)}
\end{array}\right)
=
\mathbf{M}_\rho
\left(\begin{array}{cc}
-2\lambda |\mathbf{l}|^{-1} \mathbf{H}^{(11)} + \boldsymbol{\mathcal{B}}^{(11)} - \psi\mathbf{H}^{(11)} &
\gamma_{,\phi}\mathbf{H}^{(12)} + \boldsymbol{\mathcal{B}}^{(12)} \\
-\gamma_{,\phi}\mathbf{H}^{(21)} + \boldsymbol{\mathcal{B}}^{(12)} &
-2\lambda |\mathbf{l}|^{-1} \mathbf{H}^{(22)} + \boldsymbol{\mathcal{B}}^{(22)} - \psi\mathbf{H}^{(22)}
\end{array}\right)
\left(\begin{array}{c}
\brho^{(1)} \\ \brho^{(2)}
\end{array}\right),
\end{equation}
\end{widetext}
where $\mathbf{M}_\rho$ is the mobility tensor for disconnection redistribution,
$\brho^{(k)} \equiv (\rho_1^{(k)}, \rho_2^{(k)}, \cdots)\utranspose$,
$\mathbf{H}^{(kl)} \equiv \mathbf{h}^{(k)} \otimes \mathbf{h}^{(l)}$ with $\mathbf{h}^{(k)} \equiv (h_1^{(k)}, h_2^{(k)}, \cdots)\utranspose $ and $\lambda$ is the Lagrange multiplier to ensure the normalization of the unit tangent vector $l_1^2 + l_2^2 = 1$. $\boldsymbol{\mathcal{B}}^{(kl)}$ is a nonlocal operator that encodes the stress at each point on the interface originating from all (other) disconnections in the system; application to a generic function $g(s)$ yields
\begin{widetext}
\begin{equation}\label{hatBklmn}
\mathcal{B}_{mn}^{(kl)} g(s)
\equiv
\int \rmd s_0
~~
(\mathbf{b}_m^{(k)} \times \boldsymbol{\xi})
\cdot
\left\{
\frac{\mu(1+\delta_{kl}\delta_{mn})}{2\pi(1-\nu)|\mathbf{l}|}
\left[
\ln\frac{r(s,s_0)}{r_0} \mathbf{I}
+ \hat{\mathbf{r}}(s,s_0) \otimes \hat{\mathbf{r}}(s,s_0)
\right]
\right\}
(\mathbf{b}_n^{(l)} \times \boldsymbol{\xi}) \cdot g(s_0),
\end{equation}
\end{widetext}
where $\mathbf{b}_m^{(k)}$ is the Burgers vector associated with the $m^{\rm th}$ mode disconnection on reference R$^{(k)}$, $\boldsymbol{\xi}$ is the disconnection line direction, $\mu$ and $\nu$ are the shear modulus and Poisson's ratio, $r_0$ is the dislocation core size, $r(s, s_0)$ and $\hat{\mathbf{r}}(s, s_0)$ are the magnitude and unit vector connecting points $s$ and $s_0 $.

As in dynamic optimization~\cite{kameswaran2006simultaneous}, the Lagrange multiplier $\lambda$  evolves to maintain the geometric constraint dynamically.
$\lambda$ evolves as
\begin{equation}\label{lambda}
\dot{\lambda}
= \frac{M_\lambda}{\vert \mathbf{l}\vert} \left[\left(\sum_m h_m^{(1)}\rho_m^{(1)}\right)^2 + \left(\sum_m h_m^{(2)}\rho_m^{(2)}\right)^2 - 1\right],
\end{equation}
where $M_\lambda$ is the penalty coefficient. Numerical simulations are performed by solving Eqs.~\eqref{EOM_x}, \eqref{EOM_rho} and \eqref{lambda}. Extensions to GBs/heterophase interfaces with more than two reference interfaces, as well as the numerical algorithm and definitions of the reduced quantities used in our simulations, are reported in the SM.

We first investigate the evolution of an initially circular grain (with a radius of 100) rotated with respect to the surrounding crystalline matrix  (see Fig.~\ref{fig_example}); this is a common feature observed in mazed microstructures in thin textured films~\cite{radetic2012mechanism}. The bicrystallography of the chosen GB has two orthogonal R$^{(1)}$ and R$^{(2)}$ reference interfaces with a GB energy ratio $\gamma^{(2)}/\gamma^{(1)} = 2$.
Two disconnection modes are considered on each reference interface.
We choose representative parameters $b^{(1)}$=$1$, $b^{(2)}$=$1.5$, $h_1^{(1)}$=$b^{(2)}$, $h_2^{(1)}$=$-2b^{(2)}$, $h_1^{(2)}$=$-b^{(1)}$ and $h_2^{(2)}$=$2b^{(1)}$.
We set $\rho_1^{(k)} /\rho_2^{(k)} = -1$ as an initial condition; this implies that initially the Burgers vector component normal to the GB is zero and there is no long-range stress (see $t = 0$ in Figs.~\ref{fig_example}e, f).
(All other  numerical simulation parameters are given in SM.)

The evolution of the circular GB driven by capillarity and the self-stress (from the disconnection Burgers vectors) is shown in Fig.~\ref{fig_example}a.
The grain gradually shrinks and forms a rectangular morphology with facets corresponding to the low-energy R$^{(1)}$ and R$^{(2)}$ interfaces.
The aspect ratio of the shrinking rectangular grain is  $\sim4/3$; this is a balance between the GB energy anisotropy and differences in the shear-coupling factor, $\beta_m^{(k)} \equiv b_m^{(k)}/h_m^{(k)}$.
The shrinkage rate is nearly constant until the grain size becomes very small (see the solid red line in Fig.~\ref{fig_example}c).
To track the evolution of disconnection densities $\rho_m^{(k)}$, we define the average disconnection density along the GB: $\varrho_m^{(k)} \equiv \oint \vert \rho_m^{(k)} \vert \rmd L / \oint \rmd L$.
The evolution of $\{\varrho_m^{(k)}\}$  is shown in Fig.~\ref{fig_example}d.
This evolution corresponds to the generation of net Burgers vector component perpendicular to the GB ($\rho^\perp$); see Fig.~\ref{fig_example}e for $\psi=0$.
$\rho^\perp$ is positive on the horizontal segments of the GB (the slowly migrating segments).
The increase of $|\rho^\perp|$ produces a non-uniform  stress 
(Fig.~\ref{fig_example}f for $\psi=0$).
This stress contributes to grain rotation (see the rotation angle field $\Delta\theta$  in Fig.~\ref{fig_example}g for $\psi=0$).
The grain rotates in the sense of increasing misorientation (the solid blue line in Fig.~\ref{fig_example}c represents the rotation at the grain center) with a nearly constant rotation rate (except for small grain sizes).
The variation of rotation within the grain is a manifestation of an elastic displacement gradient.

Application of an external shear stress modifies grain evolution; as can the jumps in the free energy density across the interface (as occurs in phase transformation or recrystallization).
Here, we consider a free energy density jump $\psi = -0.01$ (the free energy density within the circular grain is smaller than outside by $0.01$) leading to the growth of the embedded grain; see Fig.~\ref{fig_example}b.
The aspect ratio of the grain is reduced ($<4/3$) from the pure GB energy anisotropy result by the application of the free energy jump (an isotropic driving force).
This also influences the disconnection densities along the GB (dashed lines in Fig.~\ref{fig_example}d).
Note that the application of the free energy jump leads to negative $\rho^\perp$ (sign determined by the step height convention) during grain growth and that the stress field within the grain has the opposite sign compared with the $\psi = 0$ case (\textit{cf.} Figs.~\ref{fig_example}e, f).
Negative $\rho^\perp$  concentrates along the vertical (slow-moving) segments of the GB. We remark that the generated $\rho^\perp$ is always found along slowly moving GB segments.
Since GB migration requires the glide of disconnections with Burgers vectors perpendicular to the GB and $\rho^\perp$ is conserved,  the slow  GB segments adjust $\rho^\perp$ to accommodate the shear-induced migration of the fast segments.
In this case, the grain rotation is clockwise and decreases grain misorientation; opposite to the situation with no energy jump across the interface (\textit{cf.}  the $\psi=0$ case in Fig.~\ref{fig_example}c).

\begin{figure}[t]
\includegraphics[width=0.9\linewidth]{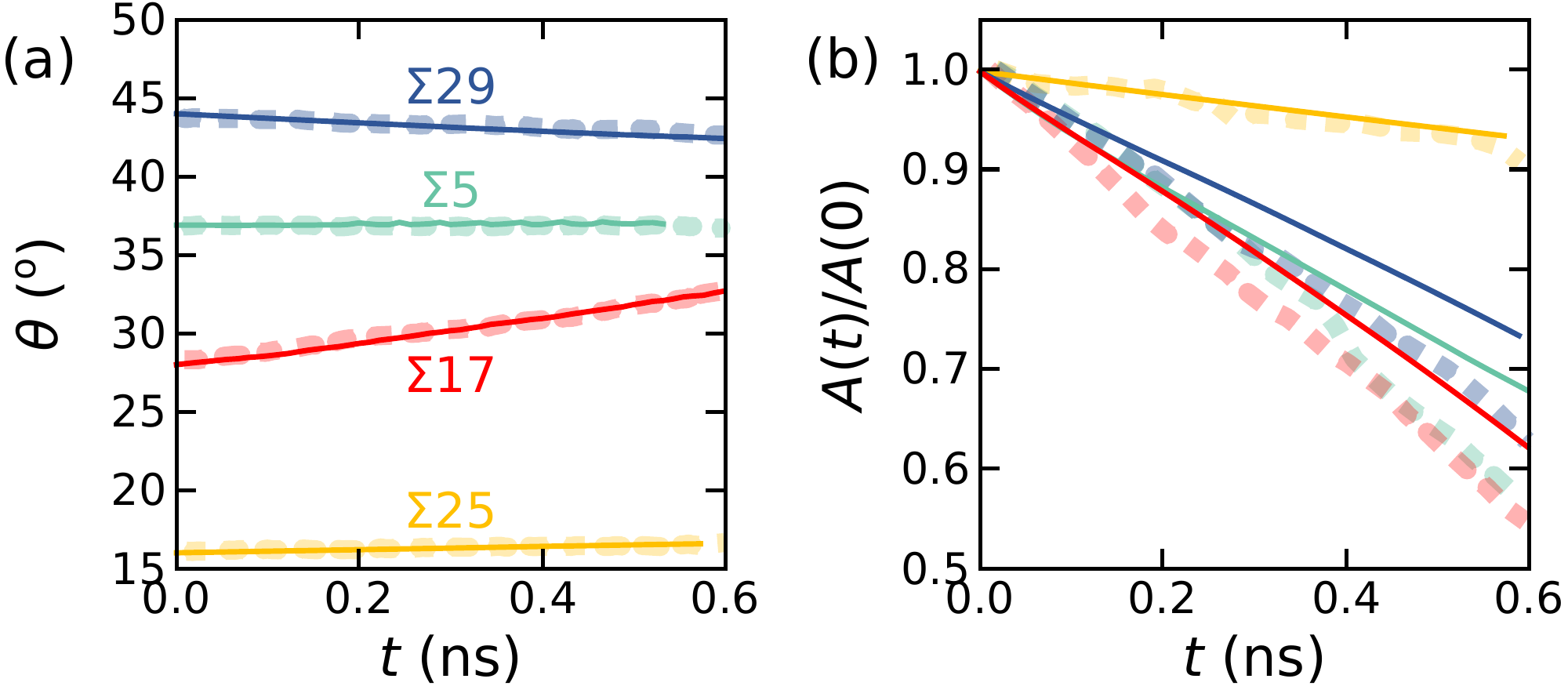}\hspace{-1.78em}%
\caption{(a) Temporal evolution of the misorientation $\theta$  and (b) reduced grain area $A(t)/A(0)$ with time $t$ for $\Sigma5, \Sigma17, \Sigma25$ and $\Sigma29 [100]$ tilt GBs from the present continuum (solid lines) and MD simulations  \cite{trautt2012grain} (dotted lines), reproduced with permission. (Copyright 2012 Elsevier)}
\label{fig_MD}
\end{figure}

We now study several cases previously examined through MD simulations~\cite{trautt2012grain}.
A circular grain was embedded in an FCC Cu matrix with rotations of $\theta = 16.26^\circ$ ($\Sigma 25$), $28.07^\circ$ ($\Sigma 17$), $36.87^\circ$ ($\Sigma 5$) or $43.60^\circ$ ($\Sigma 29$) about a $[100]$ axis.
For these misorientations, we consider four reference interfaces with their disconnections identified from bicrystallography (see SM); e.g., the reference interfaces for the $\Sigma 5$ misorientation are $(013)$, $(0\bar{3}1)$, $(0\bar{1}2)$ and $(0\bar{2}\bar{1})$.
Simulations were performed at 800~K based on our continuum model (see SM for simulation details).
Most parameters are physically based; the only parameter determined from fitting is $M_\rho / M_x$.
Our numerical simulations show that $M_\rho / M_x$ influences the grain rotation rate but not the rotation direction.
Comparing with earlier single-disconnection-mode work ~\cite{zhang2017equation,han2022disconnection} (see SM for details), we find that, considering only one disconnection mode, $M_\rho / M_x$ is proportional to $h^2\beta$.
It is also confirmed that changing the value of $M_\rho / M_x$ does not affect the sign of the shear coupling and, thus, the rotation direction.
We fit the value of $M_\rho / M_x$ for each case such that the rotation rates from our model reproduce the MD results (solid and dashed curves in Fig.~\ref{fig_MD}a, respectively).
In Figure~\ref{fig_MD}b, we compare the grain shrinkage rates from our model and those from MD simulations.
We find that the circular grain shrinkage rates vary with bicrystallography such that $\Sigma 17>\Sigma 5>\Sigma 29 \gg \Sigma 25$; consistent with  MD results (deviations $\leq15\%$).
These results demonstrate that our continuum model correctly describes the basic microscopic grain rotation mechanisms.

\begin{figure}[t]
\includegraphics[width=0.9\linewidth]{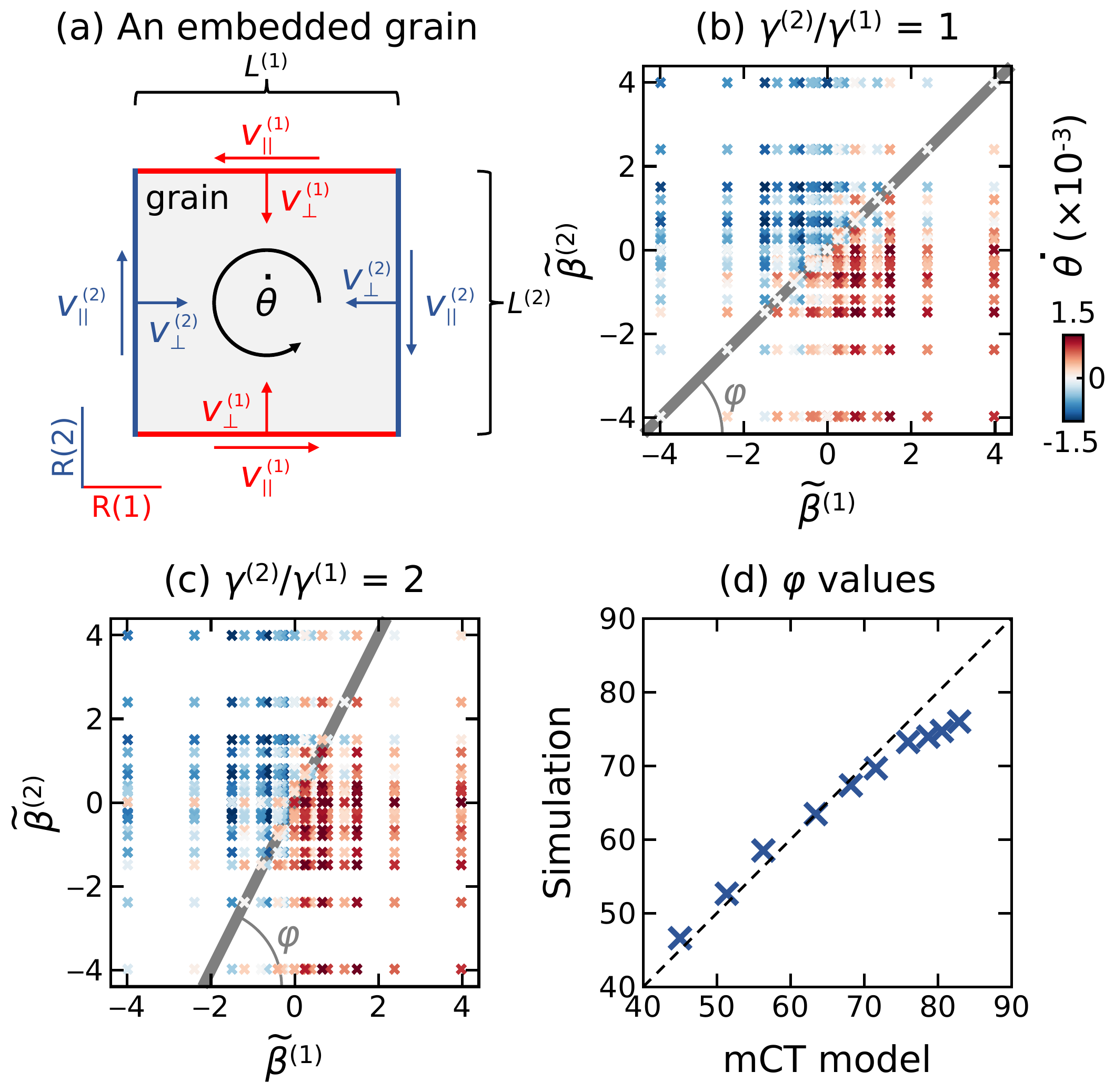}\hspace{-1.78em}%
\caption{(a) Illustration of a rectangular grain, GB velocities, and reference interfaces. Grain rotation rates $\dot{\theta}$ (indicated by color) as a function of effective shear-coupling factor $\tilde{\beta}^{(1)}$-$\tilde{\beta}^{(2)}$  for  (b) $\gamma^{(2)}/\gamma^{(1)} = 1$ and (c) $\gamma^{(2)}/\gamma^{(1)} = 2$. The dashed line represents the boundary between clockwise and counter-clockwise ratios. (d) The fitted orientation angle of the separation boundary $\varphi$  and its expected values from the modified Cahn-Taylor, mCT, model.
}
\label{fig_phase_diagram}
\end{figure}

While the model formulation may appear complex, it is possible to interpret the numerical results phenomenologically.
We extend the phenomenological Cahn-Taylor (CT) model~\cite{cahn2004unified} to include energetic and kinetic anisotropy (naming as the modified CT model, mCT).
As illustrated in Fig.~\ref{fig_phase_diagram}a, we assume that there are two reference interfaces, R$^{(1)}$ and R$^{(2)}$, and the embedded grain is rectangular with side lengths  $L^{(1)}$ and $L^{(2)}$.
Denote the steady-state migration  and sliding rates of R$^{(k)}$ interface as $v_\perp^{(k)}$ and $v_\parallel^{(k)}$; these  rates are coupled as $v_\parallel^{(k)} = \tilde{\beta}^{(k)} v_\perp^{(k)}$, where $\tilde{\beta}^{(k)}$ is the effective shear-coupling factor.
If the grain remains rectangular  while shrinking, $v_\perp^{(k)} = \dot{L}^{(l)} = M^{(k)} \partial E/\partial L^{(l)}$, where $(kl) = (12)$ or $(21)$, $M^{(k)}$ is the mobility of R$^{(k)}$, and $E = 2\gamma^{(1)}L^{(1)} + 2\gamma^{(2)}L^{(2)}$ is the total GB energy.
The rotation rate is the skew part of the velocity gradient:
\begin{equation}\label{modified_Cahn}
\dot{\theta}
= \frac{1}{2}\left(\frac{v^{(1)}_\parallel}{L^{(2)}}
- \frac{v^{(2)}_\parallel}{L^{(1)}}\right)
= \frac{M^{(1)}\gamma^{(2)}}{L^{(2)}}\tilde{\beta}^{(1)}
- \frac{M^{(2)}\gamma^{(1)}}{L^{(1)}}\tilde{\beta}^{(2)}.
\end{equation}
The effective shear-coupling factor $\tilde{\beta}^{(k)}$ is a consequence of the activation of two disconnection modes on R$^{(k)}$.
We assume that $\tilde{\beta}^{(k)}$ is a harmonic mean of the shear-coupling factors of two disconnection modes, i.e.,
$2/\tilde{\beta}^{(k)} = 1/\beta_1^{(k)} + 1/\beta_2^{(k)} = h_1^{(k)}/b_1^{(k)} + h_2^{(k)}/b_2^{(k)}$.
This can be understood as an arithmetic mean of the two-step heights weighted by the inverse of Burgers vectors; this is reasonable since the capillarity force directly acts on the step character (see the SM).

We perform numerical simulations based on our continuum model to validate the mCT model.
We set all intrinsic mobilities to 1,  $\gamma^{(2)}/\gamma^{(1)} = 1$, $b^{(1)} = b^{(2)} = 1$ and vary the step heights $\{h_m^{(k)}\}$ to sample the shear-coupling factors $\{\beta_m^{(k)}\}$ ($m = 1, 2$ and $k = 1, 2$) in the range $[-1,1]$; we consider $11^4$ cases.
We also initially set $\rho_1^{(k)}/\rho_2^{(k)} = -1$ to avoid long-range stresses in the initial configuration and then extract rotation rates; the results are displayed in the $\tilde{\beta}^{(1)}$-$\tilde{\beta}^{(2)}$ space in  Fig.~\ref{fig_phase_diagram}b.

The general trend is that the grain rotates clockwisely when $\tilde{\beta}^{(2)} > \tilde{\beta}^{(1)}$ and vice versa.
The  boundary separating the clockwise and counter-clockwise rotations is along $\tilde{\beta}^{(2)} = \tilde{\beta}^{(1)}$, consistent with the analytic solution of the mCT model.
The analytical mCT model also predicts that the configuration with the largest $|\tilde{\beta}^{(2)} - \tilde{\beta}^{(1)}|$ has the largest rotation rate.
However, our numerical simulation results show that this is not necessarily true. Indeed, in the mCT model, only capillarity is included, while elastic contributions such as \textit{self-stresses} among disconnections are absent. We also report the results for $\gamma^{(2)}/\gamma^{(1)} = 2$; see the results in Fig.~\ref{fig_phase_diagram}c.
The phase boundary between the clockwise and counter-clockwise rotations is nearly linear with $\tilde{\beta}^{(2)} = 2\tilde{\beta}^{(1)}$, consistent with the mCT model prediction.

To further validate the mCT model, we conducted a series of numerical simulations for different $\gamma^{(2)}/\gamma^{(1)} $. 
We characterize the linear phase boundary by its angle  $\varphi$, obtained by fitting the numerical simulation phase boundary orientation (see Fig.~\ref{fig_phase_diagram}b or c).
Figure~\ref{fig_phase_diagram}d compares  $\varphi$  from the mCT model with data from our numerical simulations.
We find that when $\gamma^{(2)}/\gamma^{(1)} < 5$, the $\varphi$ values from mCT model are quite close to our fitted $\varphi$ values.
When $\gamma^{(2)}/\gamma^{(1)} > 5$, our fitted $\varphi$ values are smaller than the mCT results (caused by the shape anisotropy effect when $L^{(2)}/L^{(1)}$ is not near 1).
For most GBs, the energies of the reference interfaces do not significantly differ from each other, i.e., $\gamma^{(2)}/\gamma^{(1)} < 5$ (except for some special GBs, such as coherent twins).
The mCT model provides a simple and accurate description of the grain rotation direction, demonstrating that this aspect is mainly affected by capillarity.
The mCT model cannot accurately predict the rotation rates because it does not include the elastic effects, which are fully accounted for in our continuum model exploited for numerical simulations.

Interface migration in crystalline materials is controlled by the motion of line defects, disconnections, constrained to move along the interface.
Disconnections have both step and dislocation character.
We proposed a continuum model to describe the migration of an arbitrarily-curved GB mediated by the motion of multi-mode disconnections, which respects the underlying bicrystallography, and numerically simulated the rotation of an embedded grain.
Disconnection densities along the GB evolve as the grain changes size and shape, generate stress fields, and contribute to grain rotation.
Our numerical simulations showed that grains might rotate to increase or decrease their misorientation, depending on both bicrystallography and external factors such as the application of external stress or differences in free energy density of the delimiting grains.
The predictions are quantitative, as validated with MD simulation results.
We modified the empirical Cahn-Taylor model for grain rotation that includes anisotropy and showed that it can predict the direction of rotation for embedded grains in the case of capillarity-driven growth.
The approach and bicrystallography-respecting model described here provides a quantitative approach to predicting microstructure evolution via the motion of all types of grain boundaries and/or heterophase interfaces.

\begin{acknowledgments}
JH and CQ acknowledge support from the Early Career Scheme (ECS) of Hong Kong RGC Grant 9048213 and Donation for Research Projects 9229061.
DJS and CQ acknowledge support from the Hong Kong Research Grants Council Collaborative Research Fund C1005-19G.
MS acknowledges the support of the Emmy Noether Programme of the German Research Foundation (DFG) under Grant SA4032/2-1.
\end{acknowledgments}

\bibliographystyle{apsrev4-2}
\bibliography{mybib}

\end{document}